\documentclass[prd, preprint, 12pt]{revtex4-1}

\usepackage{amsmath}
\usepackage{amssymb}
\usepackage{setspace}
\usepackage{graphicx}
\usepackage{natbib}
\usepackage{float}
\usepackage{tensor}
\usepackage[utf8]{inputenc}
\usepackage{amsfonts}
\usepackage{braket}
\usepackage{esint}

\begin{document}
 
 % ! TEX spellcheck
 %
%\ \vskip 1.0 in

\begin{center}
 { \large {\bf Proposal for a new quantum theory of gravity II}}\\
 {\large {\it - Spectral equation of motion for the atom of space-time-matter - }}

%\smallskip

\vskip 0.3 in

{\large{\bf Tejinder P.  Singh}}

%{\it $^{*}$Indian Institute of Technology Bombay, Powai, Mumbai 400076, India}\\  
{\it Tata Institute of Fundamental Research}\\
{\it Homi Bhabha Road, Mumbai 400005, India}

\smallskip

{\tt tpsingh@tifr.res.in} 

\end{center}

\bigskip
\bigskip

\centerline{\bf ABSTRACT}
\noindent In the first paper of this series, we have introduced the concept of an atom of space-time-matter [STM], which is described by the spectral action of non-commutative geometry,  corresponding to a classical theory of gravity. In the present work, we use the Connes time parameter, along with the spectral action, to incorporate gravity into trace dynamics. We then derive the spectral equation of motion for the STM atom, which turns out to be the Dirac equation on a non-commutative space.

\bigskip

 \bigskip

\setstretch{1.25}
\noindent 
This paper should be read as a follow-up to the first paper in this series \cite{Singh2019qg}, which will be hereafter referred to as I.

\section{Non-commutative geometry, trace dynamics, and gravity}

In I, we have introduced the concept of an atom of space-time-matter [STM], which is described by the spectral action of non-commutative geometry,  corresponding to a classical theory of gravity. 
We also introduced there the four levels of gravitational dynamics.
In the present work, we use the Connes time parameter, along with the spectral action, to incorporate gravity into trace dynamics. We then derive the spectral equation of motion for the STM atom, which turns out to be the Dirac equation on a non-commutative space.

In non-commutative geometry \cite{Connes2000}, the definition of the spectral action is motivated by the spectral definition of infinitesimal distance using the distance operator $d\hat{s}$. This operator is in turn related to the Dirac operator $D$ as $d\hat{s}=D^{-1}$, providing a new definition of distance, which is equivalent to the standard definition of distance [in terms  of the metric] as and when a Riemannian geometry and a manifold exists. More importantly, this spectral definition of distance continues to hold even when an underlying manifold is absent, as for instance when the algebra of coordinates does not commute. This is the essence of Connes' non-commutative geometry program.

In non-commutative geometry, the integral $\fint T$ of a first order infinitesimal in operator $T$ is defined as the coefficient of the logarithmic divergence in the Trace of T \cite{Connes2000}. [In a simplistic manner, the integral of an operator could be visualised as the sum of its eigenvalues]. The spectral action relating to gravity $S$ is defined as the slash integral
\begin{equation} 
S = \fint d\hat{s}^2 = \fint D^{-2}
\end{equation}
a definition which holds whether or not an underlying spacetime manifold is present.  When a manifold is present, this spectral action can be shown to be equal to the Einstein-Hilbert action, in the following manner. The non-commutative integral $\fint d\hat{s}^2 = \fint D^{-2}$ is given by the Wodzicki residue $Res_{W}D^{-2}$, which in turn is proportional to the volume integral of the second coefficient in the heat kernel expansion of $D^{2}$. The Lichnerowicz formula relates the square of the Dirac operator to the scalar curvature, thus enabling the remarkable result \cite{Landi1999}
\begin{equation}
\fint d\hat{s}^2 = - \frac{1}{48\pi^2} \int_{M} d^4x\; \sqrt{g} \ R
\end{equation}
In the context of the standard model of particle physics coupling to gravity, the spectral action of the gravity sector can be written as a simple function of the square of the Dirac operator, using a cut-off function $\chi(u)$ which vanishes for large $u$ (\cite{Landi1999} and references therein)
\begin{equation}
S_G[D] =  \kappa Tr [\chi (L_P^2 D^2)]
\label{spect}
\end{equation} 
The constant $\kappa$ is chosen so as to get the correct dimensions of action, and the right numerical coefficient. 

At curvature scales smaller than Planck curvature, this action can be related to the Einstein-Hilbert action using the heat kernel expansion:
\begin{equation}
S_G[D] = L_P^{-4} f_0 \; \kappa \int_M d^4x\ \sqrt{g} + L_p^{-2} f_2 \kappa \int_M d^4x \sqrt{g} R + ...
\label{sgd}
\end{equation}
Here, $f_0$ and $f_2$ are known functions of $\chi$ and the further terms which are of higher order in $L_p^{2}$ are ignored. Also, we will not consider the cosmological constant term for the purpose of the present discussion.

Let us compare and contrast the above definition of spectral action with how a trace action is defined in Adler's theory of trace dynamics \cite{Adler:04}. This theory is a classical matrix dynamics of matter degrees of freedom on a Minkowski space-time, in which physical observables are operators $q_{i}$ [defined piecewise at every spatial point] which satisfy arbitrary commutation relations amongst each other. The Lagrangian of the  theory is the matrix trace of an operator polynomial made from $q_i$ and $dq_i/dt$, where the time derivative is with respect to ordinary Minkowski time $t$. Thus, we can express the Lagrangian $L$ as $L = Tr [P(q_i,\dot{q}_i$)] where $P$ is a polynomial function of the operators $q_i$ and $\dot{q}_i$. The action of the theory is then given by
\begin{equation}
S = \int dt \; L = \int dt \; Tr [P(q_i,\dot{q}_i)] 
\end{equation}
A continuum spatial limit can also be taken, in which case the action is the four-volume integral of a trace Lagrangian density.

A mathematically well-defined concept of a trace derivative is introduced, which allows one to differentiate the trace of an operator polynomial with respect to an operator. Denoting the trace derivative of the trace Lagrangian by $\partial L/\partial q_i$, the variation of the action with respect to the configuration variables $q_i$ yields the standard Lagrange equations of motion:
\begin{equation}
\frac{d\; }{dt} \bigg(\frac{\partial L}{\partial {\dot{q}_i}}\bigg) - \frac{\partial L} {\partial q_i} = 0
\label{lag}
\end{equation}

The trace dynamics framework is general and in principle includes all matter fields and interactions [bosonic as well as fermionic], except gravity. If the gravitational degrees of freedom are raised to the status of operators, we run into problems if we still try to retain the trace dynamics formalism. How do we define determinant of the metric? Furthermore, we know from the Einstein hole argument that it is not possible to operationally define points on a space-time manifold if we do not have a classical (non-operator valued) metric overlying the manifold.

To overcome these problems which arise while trying to incorporate gravity into trace dynamics, we appeal to non-commutative geometry, and in particular to the spectral action in (\ref{spect}) above. This action, which includes the Einstein-Hilbert action as a part of its heat kernel expansion, is the trace of an operator, namely $D^2$. Comparing however with trace dynamics, we see the difference in the two cases: in trace dynamics it is the Lagrangian [not the action] which is made of trace of a polynomial. Thus, the way things stand, we cannot use the spectral action directly in trace dynamics to bring in gravity into matrix dynamics. We need to think of the spectral action as a Lagrangian, and we then need to integrate that Lagrangian over time, to arrive at something analogous to the action in trace dynamics. We can convert the spectral action into a quantity with dimensions of a Lagrangian, simply by multiplying it by $c/L_p$ (equivalently, dividing by Planck time). But which time parameter to integrate the Lagrangian over? The space-time coordinates have already been assumed to be non-commuting operators, especially in the definition of the atom of space-time-matter, the case that we are interested in. So it seems as if we have a Lagrangian, but we do not have a time parameter over which to integrate the Lagrangian, so as to make an action.

Fortunately, non-commutative geometry itself comes with a ready-made answer! The required time parameter is the Connes time $\tau$, which we discussed in I. In NCG, according to the 
Tomita-Takesaki theorem, there is a one-parameter group of inner automorphisms of the algebra ${\cal A}$ of the non-commuting coordinates - this serves as a `god-given' (as Connes puts it) time parameter with respect to which non-commutative spaces evolve \cite{Connes2000}. This Connes time $\tau$ has no analog in the commutative case, and we employ it here to describe evolution in trace dynamics. Thus we define the action for gravity, in trace dynamics, as
\begin{equation}
S_{GTD} =\kappa \frac{c}{L_P} \int d\tau \; Tr [\chi (L_P^2 D^2)]
\end{equation}
Note that $S_{GTD}$ has the correct dimensions, that of action.

The above action defines the gravity part of the (torsion-free) action for an atom of space-time-matter. We will return to considerations of torsion and the matter part of the action in future work. For now, our goal is to set up an action for gravity in trace dynamics. In so doing, we realise that we have to modify our proposed action at Level 0 in I, so as to now have a time integral over Connes time in the action. 

Next, we would like to derive the Lagrange equations (\ref{lag}) for this trace action. For this we need to figure out what the configuration variables $q$ are. In the presence of a manifold, those variables would simply be the metric. But we no longer have that possibility here. We notice though that the operator $D$ is like momentum, since it has dimensions of inverse length. $D^2$ is like kinetic energy, so its trace is a good candidate Lagrangian. Therefore, we define a new operator $q$, having the dimension of length, and we define a velocity $dq/d\tau$, which is defined to be related to the Dirac operator $D$ by the following new relation
\begin{equation}
D \equiv \frac{1}{Lc}\;  \frac{dq}{d\tau}
\end{equation}
where $L$ is a length scale associated with the STM atom. The action for the STM atom can now be written as
action for gravity, in trace dynamics, as
\begin{equation}
S_{GTD} =\kappa \frac{c}{L_P} \int d\tau \; Tr [\chi (L_P^2 \dot{q}^2/L^2 c^2)]
\end{equation}
where the time derivative in $\dot{q}$ now indicates derivative with respect to Connes time. This clearly is the trace action for a free `operator' particle where $q$ represents the to-be-gravitational degree of freedom in 
non-commutative geometry. Only when a background manifold is available, do the $q$-s get related to the metric. We can now vary the action of the free particle with respect to the position operator $q$, and by taking the trace derivative in the Lagrange equations of motion it is easy to conclude that the equation of motion is $\ddot{q}=0$. Thus the velocity is constant, and the Dirac operator is proportional to a constant matrix, and we can write the solution to the equations of motion as an eigenvalue equation
\begin{equation}
D\psi = \frac{1}{L}\psi
\label{deq}
\end{equation}
where the state vector depends on Connes time $\tau$, and on the gravitational degree of freedom $q$. When a manifold is available, then on scales below Planck length this equation is equivalent to Einstein equations, by virtue of the expansion (\ref{sgd}). Note that on a manifold, this Dirac operator is same as the Dirac operator on a [torsion free] curved spacetime.

This completes the description of the gravity sector of the STM atom introduced at Level 0 in I. It remains to generalise this to an asymmetric metric, and include torsion, and matter. Restricting for the moment to the gravity sector, we see that its description is very simple. Each STM atom is a free particle, and we have an ideal gas of non-interacting STM atoms at the leading order approximation. Entanglement between STM atoms brings about interaction, and it remains to be understood how and why entanglement causes spontaneous localization, leading to the emergence of classical matter fields and classical space-time. [We note that upon the emergence of a classical spacetime, Connes time is lost, and ordinary locally Lorentz invariant time is recovered]. Because we are now describing gravity in the trace dynamics framework, there is a conserved Adler-Millard charge $\tilde{C}$ and the constant $\kappa$ above can be identified with this charge [recall that the only fundamental constants at Level 0 are Planck length and speed of light]. When we perform the statistical thermodynamics of the STM atoms at Level 0, equipartition of the Adler-Millard charge occurs, resulting in the emergence of the Planck constant $\hbar$, and hence Newton's gravitational constant $G$. Canonical quantum commutation relations for the canonical pair $(q,D)$ emerge, as do the quantum Heisenberg equations of motion. It remains to be shown how spontaneous collapse caused by thermodynamic fluctuations (equivalently entanglement) at Level I gives rise to classical spacetime and classical general relativity at Level III. This work is currently in progress. 

\section{Discussion}
Non-commutative geometry was motivated by quantum theory: to carry over the non-commutativity in quantum phase space to geometry in general. This is achieved by mapping Riemannian geometry to an algebra of [commuting] functions. The algebra is then generalised to a non-commutative algebra, and said to be equivalent to a non-commutative `geometry', even though there is, strictly speaking, no geometry left. More importantly though, the relation of noncommutative geometry [NCG] to quantum gravity and physics in general has remained somewhat unclear. NCG is not needed in the classical universe, which to an excellent approximation is described by general relativity on a Riemannian manifold. Nor does it make sense to quantize an NCG: why should one quantise something which is already `quantum' in nature? If NCG is by itself a quantum theory of gravity, then what is NCG's relation to quantum theory?

However, in the context of trace dynamics, NCG appears very natural. Trace dynamics is a beautiful [classical] matrix dynamics from which one obtains quantum [field] theory as a statistical thermodynamics of the underlying matrices. But this has been done for matter fields, on a Minkowski spacetime background. Gravity was not yet incorporated. At Level 0 in our layer diagram for gravity and matter, NCG blends perfectly with trace dynamics, providing the necessary `matrix  dynamics' type description of gravitation. This, provided we take into account the existence of the natural time parameter $\tau$  in NCG. Then, the unified description of NCG and trace dynamics maybe expressed by an action of the form
\begin{equation}
S =\kappa \frac{c}{L_P} \int d\tau \; Tr [\chi (L_P^2 D^2)] + \int d\tau \; Tr [P(q_i,\dot{q}_i)] 
\end{equation}
at Level 0. Since the Adler-Millard charge $\tilde{C}$ is the only constant with dimensions of action at Level 0, we identify $\kappa$ with $\tilde{C}$.  Moreover, having introduced the time parameter $\tau$ it is more reasonable to work with Planck time $\tau_{Pl}=L_{pl}/c$ and to think of the speed of light as derived from Planck length and Planck time. So we can write the above action as
\begin{equation}
\frac{S}{\tilde{C}} = \int \frac{d\tau}{\tau_{Pl}}\;  \bigg[ Tr [\chi (L_P^2 D^2)] + \frac{\tau_{P}l}{\tilde{C}}Tr [P(q_i,\dot{q}_i\bigg]
\end{equation}
It remains to be seen how these two parts of the action should be merged together so as to arrive at the action for the STM atom. [We have made one specific suggestion in this regard, in I.]  Once that has been done, the statistical thermodynamics of the STM atoms will yield a quantum theory of gravity at Level I.

The following important aspects relating to the present work deserve further investigation:

\begin{itemize}

\item Why does the Dirac operator enter in the spectral definition of the Einstein-Hilbert action? Is there a relation between the Dirac equation and Einstein equations? Perhaps yes, if we recall that in our theory Einstein equations are supposed to arise after the spontaneous collapse of many entangled atoms, each of which obeys a Dirac equation, as derived in (\ref{deq}) above. 

\item What are the properties of the eigenvalues and corresponding eigenspinors, of the Dirac operator and its square, in the non-commutative space at Level 0?

\item At Level 0 in the gravitational dynamics, $\hbar$ does not appear. The only quantity with dimensions of spin is the Adler-Millard charge, which refers to the system as a whole, not to individual STM atoms. Thus it seems to us that one cannot assign a quantum spin to an STM atom at Level 0. Spin, as well as mass, are concepts emergent only at Level I. At Level 0, there is only a length scale associated with every STM atom, and perhaps one must not make a distinction between spin and mass. The fundamental constants at Level 0 are Planck length and Planck time, and the conserved quantity is the Adler-Millard charge. Planck mass $m_{Pl} = (\hbar c/G)^{1/2}$, along with $\hbar$ and $G$, is emergent only at Level I. 

\item The `inner automorphisms' of NCG that are responsible for the existence of Connes time, and the global unitary invariance of the matrix dynamics responsible for the conserved Adler-Millard charge, are 
possibly both one and the same symmetry. After all, NCG is a matrix dynamics of gravity, as we have seen in this work. It is possible that the Adler-Millard charge arises as a Noether conserved charge, corresponding to the invariance of the theory under translations in Connes time $\tau$. This would reinforce our assertion that NCG is the matrix dynamics of gravity, in the sense of trace dynamics.

\item Given the spectral action for the STM atom at Level 0, what is the reason for spontaneous collapse of entangled STM atoms? For collapse to take place, it seems essential to make the metric asymmetric, and introduce torsion. This ensures  that there is a non-unitary component in the evolution, brought about by the anti-symmetric part of the metric. Nonetheless, norm of the state vector continues to be preserved in Connes time, because the evolution is geodesic.

\end{itemize}

\bigskip

\bigskip

\noindent Acknowledgements: For valuable discussions and ongoing collaboration I would like to thank Harish M.,
Maithresh Palemkota, 
B J Arav, 
Shivnag Sista, 
Diya Kamnani,
Guru Kalyan, 
Shubham Kadian, 
Siddhant Kumar, 
Ankur Rohilla, 
Abhinav Varma,
Shounak De, 
Shreyansh Singh, 
Ish Mohan Gupta and 
Krishnanand K Nair.

\bigskip

\centerline{\bf REFERENCES}

\bibliography{biblioqmtstorsion}

\end{document}